\documentclass[12pt]{iopart}
\usepackage{graphicx}
\usepackage{color}
\usepackage{amssymb}
\usepackage{amsthm}
\begin{document}

\title[Sequence heterogeneity and the dynamics of molecular motors]{Sequence heterogeneity and the dynamics of molecular motors}

\author{Yariv Kafri$^*$\dag\ and David R. Nelson\ddag}

\address{\dag\ Physicochimie Curie (CNRS-UMR168), Institut Curie,
Section de Recherche, 26 rue d'Ulm 75248 Paris Cedex 05, France}

\address{\ddag\ Department of Physics, Harvard University,
Cambridge, MA 02138}

\begin{abstract}
The effect of sequence heterogeneity on the dynamics of molecular
motors is reviewed and analyzed using a set of recently introduced
lattice models. First, we review results for the influence of
heterogenous tracks such as a single-strand of DNA or RNA on the
dynamics of the motors. We stress how the predicted behavior might
be observed experimentally in anomalous drift and diffusion of
motors over a wide range of parameters near the stall force and
discuss the extreme limit of strongly biased motors with one-way
hopping. We then consider the dynamics in an environment containing
a variety of different fuels which supply chemical energy for the
motor motion, either on a heterogeneous or on a periodic track. The
results for motion along a periodic track are relevant to kinesin
motors in a solution with a mixture of different nucleotide
triphosphate fuel sources.
\end{abstract}

\maketitle

\section{Introduction}

The study of molecular motors has been transformed in recent years
with the increasing use of single molecule experiments
\cite{Bustamante2003,BusRev}. In one key experiment an external
force is applied to a molecular motor opposing its motion
\cite{Visscher99,Davenport00,Wang98,Perkins2003}. Typically, as the
force is increased, the velocity of the motor decreases until it is
completely stalled. The behavior of the velocity as a function of
force provides much information on the chemical cycle underlying the
motion of the motor. For example, the stall force is a direct
estimate of the force exerted by the molecular motor. The
experiments have also motivated much theoretical work on the
dynamics of the motors \cite{Julicher97,Ajdari,Prost,Prost94}.
Fitting the experimentally obtained velocity-force curves allows
extraction of detailed information on the chemical cycle of the
motor \cite{Fisher99,FishKol00,Kolo00}.

Most theoretical studies have focused on motors which move on
featureless, or periodic linear tracks
\cite{Julicher97,Ajdari,Prost,Fisher99,FishKol00,Kolo00}. Such a
description would be appropriate, for example, for kinesin which
moves along a microtubule filament, which is periodic, using only
ATP for its motion \cite{HowardBook,GelKin}. However, in many cases
the assumption of a periodic medium fails. Examples of motors which
move on heterogeneous tracks include RNA polymerase
\cite{Davenport00,Wang98} which moves along DNA, ribosomes which
move along mRNA, helicases \cite{Ha2002,BiancoHel} which unwind DNA,
exonucleases \cite{Xie,blockexo} which turn double-stranded DNA into
single-stranded DNA and many others. All these motors move along
tracks which are inherently ``disordered'' or heterogeneous due to
the underlying sequence of the linear template. Theoretically,
molecular motors moving along disordered tracks have received much
less attention \cite{Harms97,Kafri04,KLN05}. Another form of
heterogeneity which has largely been ignored arises from the
different chemical fuels which may be used by molecular motors to
move along the track. For example, RNA polymerase uses different
nucleotide triphosphates (NTP's) which build the mRNA it produces,
each supplying a different amount of chemical energy, to move along
a DNA strand. A different ``annealed'' form of disorder (in contrast
to the ``quenched'' disorder embodied in a particular nucleotide
sequence) can be present even in molecular motors moving along
perfectly periodic tracks in a solution containing several distinct
types of chemical fuels. For example, it is known that kinesin can
move using other nucleotide triphosphates (such as GTP) instead of
ATP, albeit less efficiently \cite{Kin1,Kin2,Kin3}.

Recently, we have introduced a simplified model for molecular motors
which allows the effects of disorder to be studied in considerable
detail \cite{Kafri04}. We have focused so far on heterogeneous
tracks and argued that near the stall force the dynamics of the
motor is strongly affected by the heterogeneity embodied in a
particular DNA or RNA sequence. Due to the ``sequence disorder'' on
which the motor is moving (many DNA sequences have only short range
correlations \cite{stanley}) the displacement of the motor as a
function of time ceases to be linear in time close enough to the
stall force. The displacement becomes {\it sublinear} in time,
growing as $t^\mu$, with $\mu$ varying continuously from $1$ to $0$
as the stall force is approached. As discussed below there are also
anomalies in the diffusive spreading about the average motor
position which extend even further below the stall force.

In this paper we review some of these results, stressing several
experiments which could be performed to test the predictions of the
model. We also explore the effect of heterogeneous fuels on the
motion of molecular motors. In \cite{Kafri04} it was suggested that
inhomogeneous fuel concentrations could enhance significantly the
regime near the stall force over which anomalous dynamics is
observed. Here we study this type of disorder numerically and
illustrate the dramatic effect of varying the concentrations of the
different fuels used to power the motor. For motors moving along
heterogeneous tracks we also discuss the dynamics in the extreme
limit where detailed balance is violated and motors {\it never} take
backward steps. Finally, we consider motors moving along a {\it
periodic} substrate powered by different kinds of fuels. We discuss,
for simple cases, the expected velocity of the motor as the relative
proportion of two different types of fuel in the solution is varied.
The behavior of more complicated models is also discussed.

\section{The Model}
\label{secmodel}

\begin{figure}
\centering
\includegraphics[width=11cm]{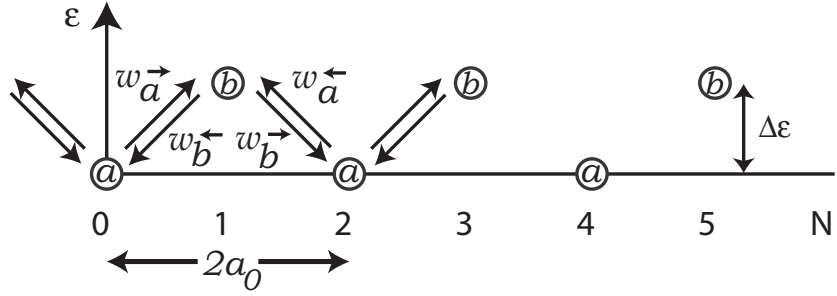} \caption{Graphical representation
of a simplified lattice model for  molecular motors and the relevant
rates
$w_a^\rightarrow,w_a^\leftarrow,w_b^\rightarrow,w_b^\leftarrow$ and
energy difference $\Delta \varepsilon$. The distinct even and odd
sublattices are denoted by $a$ and $b$ respectively.
\label{fig:motor}}
\end{figure}

In this section we define the model used throughout the paper. We
start with a special case of the general class of $n$-state models
explored by Kolomeisky and Fisher
\cite{Fisher99,FishKol00,Kolo00,KolWid} and consider a ``minimal''
motor with only two internal states. This simplified model
reproduces important features of previously studied systems and
allows us to explore generic behavior in new situations in a minimal
form. The model is easily generalized to account for heterogeneous
fuels and tracks. When appropriate, we will mention how results are
modified for general $n$-state models. The model has been introduced
and studied in detail in \cite{Kafri04} and here we only review its
basic properties. We begin by assuming a perfectly periodic
substrate. The location along the one-dimensional track, $x$, is
assumed to take a discrete set of values $x_m$, where $m=0,1,2
\ldots$ labels distinct $a$ and $b$ sites. Although not essential,
we assume for simplicity that the distances between $x_{m+1}-x_m$
and $x_{m+2}-x_{m+1}$ are equal and set $x_{m+2}-x_{m}=2a_0$, which
is the size of a step taken by the motor after completing a chemical
cycle such as hydrolysis of ATP. In general, as discussed in
\cite{Fisher99}, the distance traveled by the motor between internal
states may be different for different internal transitions. However,
we do not expect such modifications to affect the long time behavior
over a range of parameters near the stall force. The dynamics
embodied in the model is shown schematically in Fig.
\ref{fig:motor}. Internal states labeled by $a$ have an energy
$\varepsilon=0$ while internal states labeled by $b$ have a higher
energy $\varepsilon=\Delta \varepsilon$. The local detailed balance
condition (in temperature units such that $k_B=1$) is satisfied by
our choice of rate constants,
\begin{eqnarray}
w_a^\rightarrow&=&(\alpha e^{\Delta \mu /T} + \omega)e^{-\Delta
\varepsilon/T-f/2T}
 \nonumber \\
w_b^\leftarrow&=&(\alpha  + \omega)e^{f/2T}
 \nonumber \\
w_a^\leftarrow &=&(\alpha' e^{\Delta \mu /T} + \omega')e^{-\Delta
\varepsilon/T+f/2T} \label{eq:rates}
\\
w_b^\rightarrow&=&(\alpha'  + \omega')e^{-f/2T} \;.  \nonumber
\end{eqnarray}
Following Ref. \cite{Julicher97}, there are two parallel channels
for the motion. The first, represented by contributions containing
$\alpha$ and $\alpha'$, arise from utilization of chemical energy
biased by a chemical potential difference $\Delta \mu$ between, say
ATP and the products of hydrolysis ADP and P$_i$. The second
channel, represented by the terms containing $\omega$ and $\omega'$,
correspond to thermal transitions unassisted by the chemical energy.
We assume that the externally applied force $F$ biases the motion in
a particularly simple way (consistent with detailed balance) and
define $f=Fa_0$. If the substrate lacks inversion symmetry (a
necessary condition for directed motion, driven by $\Delta \mu$,
when $f=0$ \cite{Julicher97}), we have $\alpha' \neq \alpha$ and
$\omega' \neq \omega$. If the fuel is ATP, the chemical potential
difference which drives the motion is \cite{HowardBook}
\begin{equation}
\Delta \mu =T \left[\ln \left( \frac{[ATP]}{[ADP][P_i]}\right) -\ln
\left( \frac{[ATP]_{\rm eq}}{[ADP]_{\rm eq}[P_i]_{\rm eq}} \right)
\right] \;, \label{eq:dmu}
\end{equation}
where the square brackets $[ \ldots ]$ denote concentrations under
experimental conditions and the brackets $[ \ldots ]_{\rm eq}$
denote the corresponding concentrations at equilibrium.

The rate constants in Eq. \ref{eq:rates} define a set of
differential equations for the probability $P_n(t)$ of being at site
$n$ at time $t$. For odd $n$ one has
\begin{equation} \frac{dP_n(t)}{dt}=
w_a^\rightarrow P_{n-1}(t)+ w_a^\leftarrow P_{n+1}(t) -
(w_b^\rightarrow+w_b^\leftarrow) P_{n}(t) \;, \label{eq:b}
\end{equation}
while for even $n$
\begin{equation}
\frac{dP_n(t)}{dt}= w_b^\rightarrow P_{n-1}(t)+ w_b^\leftarrow
P_{n+1}(t) - (w_a^\rightarrow+w_a^\leftarrow) P_{n}(t) \;.
\label{eq:a}
\end{equation}
It is illuminating, especially when we consider rate constants which
depend on the position along a heterogenous track, to study two
limits of these equations. In the first the chemical potential
difference $\Delta \mu$ and the applied force $f$ are small compared
to the energy difference $\Delta \varepsilon$ so that $b$ states
relax quickly compared to $a$ states. This condition implies that
$(w_b^\rightarrow + w_b^\leftarrow) \gg (w_a^\rightarrow +
w_a^\leftarrow)$ so that in the long time-limit to a good
approximation the left hand side of Eq. \ref{eq:b} may be set to
zero. Upon solving for $P_n(t)$ with $n$ odd and substituting into
Eq. \ref{eq:a}, we obtain differential equations just for the {\it
even} sites
\begin{eqnarray}
\frac{dP_n(t)}{dt}&=&\frac{w_b^\leftarrow w_a^\leftarrow P_{n+2}(t)+
w_b^\rightarrow w_a^\rightarrow P_{n-2}(t)-(w_b^\rightarrow
w_a^\rightarrow+w_b^\leftarrow
w_a^\leftarrow)P_n(t)}{(w_b^\rightarrow + w_b^\leftarrow)} \;,
\nonumber\\
&& {\rm (n \; even)} \;.\label{eq:lim1}
\end{eqnarray}
Similarly in the limit $\Delta \mu \gg \Delta \varepsilon$ (with $f$
near the stall force) the motor spends most if its time in $b$
states. Now, in the long time-limit to a good approximation the left
hand side of Eq. \ref{eq:a} may be set to zero. The remaining
differential equations for the odd sites read
\begin{eqnarray}
\frac{dP_n(t)}{dt}&=&\frac{w_b^\leftarrow w_a^\leftarrow P_{n+2}(t)+
w_b^\rightarrow w_a^\rightarrow P_{n-2}(t)-(w_b^\rightarrow
w_a^\rightarrow+w_b^\leftarrow
w_a^\leftarrow)P_n(t)}{(w_a^\rightarrow + w_a^\leftarrow)} \;,
\nonumber\\
&& {\rm (n \; odd)} \;. \label{eq:lim2}
\end{eqnarray}

Note that in both limits the dynamics of the motors on long-times
can be described by a random walker moving on an effective energy
landscape associated with what is in general a non-equilibrium
dynamics. Upon absorbing the denominator factors $(w_a^\rightarrow +
w_a^\leftarrow)$ and $(w_b^\rightarrow + w_b^\leftarrow)$ into a
rescaling of the rate constants in the numerator, the effective
energy landscape can be read off from Eq. \ref{eq:lim1} or Eq.
\ref{eq:lim2}. One finds that the effective energy difference
between two sites which are two monomers apart is given by
\begin{equation}
E_{n+2}-E_{n} \equiv \Delta E= T\ln\left( \frac{w_a^\leftarrow
w_b^\leftarrow}{w_a^\rightarrow w_b^\rightarrow}\right) \;.
\label{eq:dE}
\end{equation}
For a {\it periodic} track, this leads to a tilted energy landscape
(with tilt controlled by $\Delta \mu$ and $f$) and an effective
energy difference between $2m$-adjacent monomers $2m\Delta E $. The
tilted energy landscape leads to diffusion with drift on long
time-scales and large length-scales. The effective energy landscape
can also be obtained by assuming detailed balance and equating the
rate asymmetry between two neighboring even sites to an effective
energy difference $\Delta E$. It is straightforward to verify from
Eq. \ref{eq:dE} with Eq. \ref{eq:rates} that for a periodic
substrate no net motion is generated when the external force $f=0$
and the chemical potential difference $\Delta \mu =0$. Also, when
there is directional symmetry in the transition rates,
$\alpha=\alpha'$, $\omega=\omega'$, and $f=0$ no net motion is
generated even when $\Delta \mu \neq 0$. Absent this symmetry,
chemical energy can be converted to motion. These conditions are
equivalent to those presented in \cite{Julicher97,Prost94} for
continuum models and are exhibited here in a minimal model. The
effect of the externally applied force is simply to bias the motion
of the motor in the direction in which it is applied.

The velocity for a motor moving along a {\it periodic} track in the
two limits discussed above can be obtained by taking the continuum
limit and yields $v=(w_a^\rightarrow w_b^\rightarrow- w_a^\leftarrow
w_b^\leftarrow)/(w_b^\rightarrow+ w_b^\leftarrow)$ for the limit
specified by Eq. \ref{eq:lim1} and $v=(w_a^\rightarrow
w_b^\rightarrow- w_a^\leftarrow w_b^\leftarrow)/(w_a^\rightarrow+
w_a^\leftarrow)$ for the limit specified by Eq. \ref{eq:lim2}. More
generally, for periodic rates, it is straightforward to calcualte
the velocity, for example using Bloch eigenfunctions $|k \rangle
\sim e^{ikx}$ and expanding the eigenvalues in the wavevector $k$.
The linear term gives the velocity and the quadratic part the
effective diffusion constant. For the velocity one finds
\begin{equation}
\label{eq:vel} v=\frac{w_a^\rightarrow w_b^\rightarrow-
w_a^\leftarrow w_b^\leftarrow}{w_a^\rightarrow+ w_b^\rightarrow+
w_a^\leftarrow+ w_b^\leftarrow} \; .
\end{equation}

An alternative to studying the solution of the equations, which will
be very useful throughout this paper, is to use Monte-Carlo
simulations. The rates specified in Eq. (\ref{eq:rates}) can be
simulated using the following procedure: To make the simulation
efficient we first normalize the entering or leaving rates for a
site so that the largest one is unity. Then, at each step we choose
with equal probability attempting to move the motor to the right or
left on the lattice. Following this choice a random number is drawn
from a uniform distribution in the interval $[0,1]$. The motor is
moved in the chosen direction provided the random number is smaller
than the corresponding rate. Thus, if a motor finds itself on site
$a$ in Fig. 1 and
$w_a^\rightarrow>w_a^\leftarrow,w_b^\rightarrow,w_b^\leftarrow$ is
the largest rate it will (after the rescaling) move one step to the
right with probability $1/2$, one step to the left with probability
$1/2\left(w_a^\leftarrow / w_a^\rightarrow \right)$ and it will stay
put with probability $1/2[ 1-(w_a^\leftarrow / w_a^\rightarrow)]$.
Note that the probability of actually moving one step (right or
left) to the $b$-sublattice during a particular attempt is $1/2[
1+(w_a^\leftarrow / w_a^\rightarrow)]$. Once the $b$-sublattice is
reached the procedure is repeated with the rates $w_b^\rightarrow$
and $w_b^\leftarrow$ (note that the probabilities are still obtained
by dividing by $w_a^\rightarrow$). To compare, for example,
velocities for different choices of rates the overall number of
attempts is rescaled at the end by the fastest rate (taken to be
$w_a^\rightarrow$ in the above example). The same procedure is
followed for both homogeneous and heterogeneous tracks. For the
latter the largest rate is chosen from all possible hopping rates
along the track. This protocol ensures relaxation to equilibrium in
the absence of chemical or mechanics driving forces \cite{Newman}.

In \cite{Kafri04} we have analyzed in detail the motion of the model
when the track is not periodic. Such tracks arise naturally, for
example, for motors such as RNA polymerase or helicases which move
on DNA which has a well defined sequence. In this case the energy
difference $\Delta E (m)$ now becomes an explicit function of the
location $m$ along the track, due to the dependance of the rates on
the location on the track.

\begin{figure}
\centering
\includegraphics[width=10cm]{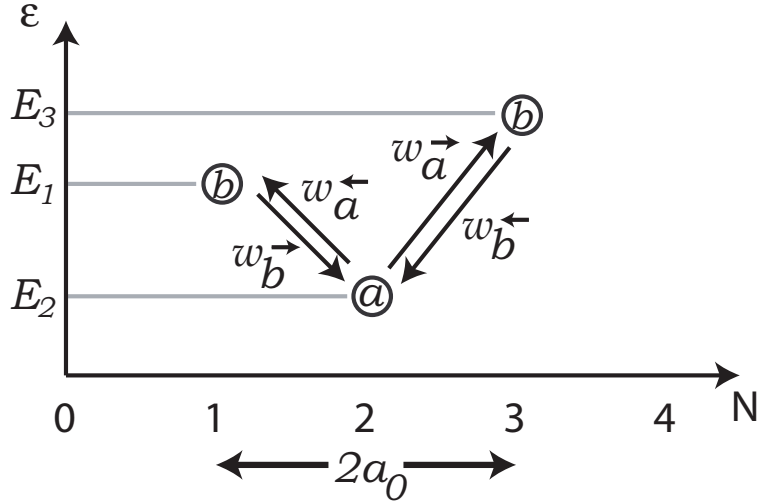} \caption{The transition
rates and energy levels of three sites, spanning two monomers
corresponding to a motor moving along a heterogenous track. Note
that $E_1 \neq E_3$ due to the non-periodicity of the track. The
transition rates now depend explicitly on the location along the
track. To avoid cluttering this dependence was suppressed in the
figure.\label{fig:three}}
\end{figure}

To understand the energy which arises for {\it heterogeneous} tracks
consider the ``integrating out'' procedure applied to the three
sites shown in Fig. \ref{fig:three}, where three distinct motor
binding energies, $E_1$, $E_2$ and $E_3$, are indicated explicitly.
We work in the limit $\Delta \mu \gg \Delta E$, with $\Delta
E=E_1-E_2$ or $\Delta E=E_3-E_2$, and $f$ close to the stall force,
so that the approximation leading to Eq. \ref{eq:lim2}
(``integrating out'' site $2$) is appropriate. As rates for the
heterogeneous cluster shown in Fig. \ref{fig:three}, we take
\begin{eqnarray}
w_a^\rightarrow (13)&=&[\alpha(13) e^{\Delta \mu(13) /T} +
\omega(13)]e^{-(E_2-E_1)/T-f/2T}
 \nonumber \\
w_b^\leftarrow (13)&=&[\alpha(13)  + \omega(13)]e^{f/2T}
 \nonumber \\
w_a^\leftarrow (13)&=&[\alpha'(13) e^{\Delta \mu(13) /T} +
\omega'(13)]e^{-(E_2-E_3)/T+f/2T} \label{eq:rateshet}
\\
w_b^\rightarrow (13)&=&[\alpha'(13)  + \omega'(13)]e^{-f/2T} \;.
\nonumber
\end{eqnarray}
where the arguments ``$(13)$'' appended to the $w$'s,
$\alpha$'s,$\alpha'$'s, $\omega$'s and $\omega'$'s simply mean that
these are the heterogenous rates appropriate to the cluster $1-2-3$.
These rates obey detailed balance conditions for the two channels,
and have a similar dependence on $\Delta \mu (13)$ and $f$ and
various energy differences as the rates in Eq. \ref{eq:rates}. The
notation $\Delta \mu (13)$ indicates that the chemical potential
difference could depend on which NTP (in the case of RNA polymerase)
provides the energy for that particular step. Upon assuming fast
relaxation of site $2$ in Fig. \ref{fig:three}, from a formula
similar to Eq. \ref{eq:dE},
\begin{eqnarray}
\Delta E_{13}&=&E_3-E_1+2f-T \ln \left[ \frac{(\alpha(13)e^{\Delta
\mu/T}+\omega(13))(\alpha'(13)+\omega'(13))}{(\alpha'(13)e^{\Delta
\mu/T}+\omega'(13))(\alpha(13)+\omega(13))} \right] \;, \nonumber \\
&\equiv& E_3-E_1+2f+\eta_{13} \label{eq:dEhet}
\end{eqnarray}
Eq. \ref{eq:dEhet} illustrates the following important points,
applicable to motor molecules on heterogenous tracks more generally:
($a$) if $\Delta \mu(13)=0$, then $E_{13}=E_3-E_1+2f$, with a
similar formula for all neighboring pairs of odd sites. Thus, in the
absence of chemical energy, we have a ``random energy landscape''
with bounded energy fluctuations; ($b$) if $\alpha(13)=\alpha'(13)$
and $\omega(13)=\omega'(13)$ (inversion symmetry), chemical energy
does not lead to net motion between sites $1$ and $3$, as discussed
above; ($c$) in general, $\Delta E_{13}=E_3-E_1+2f+\eta_{13}$, where
$\eta_{13}$ is a random function of position along the heterogeneous
track. Upon passing to a coarse-grained position $\eta(m)$, where
$m$ is the position along the track we see that the effective
`coarse grained' energy difference between two points $m_1$ and
$m_2=m_1+m$ which are $m$-monomers apart is given by
$\sum_{m'=m_1}^{m_1+m} \eta(m')$. The landscape itself behaves like
a random walk with fluctuations which grow as $\sqrt{m}$,
corresponding to a {\it random-forcing} energy landscape (for RNA
polymerase, the different chemical potentials of the nucleotides in
the transcript also contribute to a random forcing landscape
\cite{Kafri04}).

The self-similar structure of the random force landscape leads to
interesting dynamics near the stall force. As the stall force is
approached the dynamics slows down and becomes dominated by motion
between deep minima of the energy landscape \cite{Bouchaud90}. The
minima correspond to specific locations along the track where the
motor tends to pause. The distribution of dwell times at these
minima, $P(\tau)$, averaged over the different locations on the
track, is expected to behave as $\tau^{-(1+\mu)}$, where $\mu$ (not
to be confused with a chemical potential!) is related to the force,
fluctuations in the effective energy landscape and temperature. For
random forcing energy landscape where the energy difference between
two points is drawn from a {\it Gaussian} distribution with a
variance $V=\overline{\eta(m)^2}$, where the overline denotes an
average along the sequence, one can show that \cite{Bouchaud90}
\begin{equation}
\mu(f)=2T \vert \overline{\Delta E}_{f=0}-2f \vert / V \;,
\label{eq:mu}
\end{equation}
where $\overline{\Delta E}_{f=0}$ is the mean slope of the potential
(averaged along the sequence) at zero force. The exponent $\mu$ thus
decreases continuously to zero as $f$ increases toward the stall
force of the motor (defined by $\mu(f_s) \equiv 0$). For more
general distributions of the effective energy difference the value
of $\mu$ might be different from Eq. \ref{eq:mu} by factors of order
unity.

Near the stall force the expected distribution of pause times
becomes broader as $\mu$ becomes closer to $0$. The dynamics of the
motors are altered from diffusion with drift when the pause-time
distribution becomes very broad. The dynamics then depends on the
numerical value of $\mu$ defined in Eq. \ref{eq:mu}
\cite{Kafri04,KLN05,Bouchaud90}:
\begin{itemize}
\item ${\mu<1}$ -- Around the stall force, both the drift and diffusive behavior
of the motor become anomalous. The displacement of the motor as a
function of time increases as $t^\mu$. Thus, in this region the
velocity is undefined, in the sense that it depends on the
experimental observation time, $t_E$, through $v \sim t_E^{\mu-1}$.
Moreover, the spread of the probability distribution of the motor
about its mean position also behaves anomalously with a variance
which grows as $t_E^{2\mu}$. Experimentally, for a given $t_E$, this
anomaly should lead to a convex velocity as a function of force
curve in the vicinity of the stall force. The curve will become more
and more convex as $t_E$ is increased; the velocity actually {\it
vanishes} for a range of $f$'s near the stall force in the limit
$t_E \to \infty$.

\item $1<\mu<2$ -- Further away from the stall force the displacement of the motor as a function of time
grows linearly. At long-times the velocity becomes independent of
the averaging window. However, the variance of the probability
distribution around the mean is anomalous and grows as
$t_E^{2/\mu}$.

\item $\mu>2$ -- Far below the stall force both the displacement and the variance of
the probability distribution around the mean grow linearly in time,
as in conventional diffusion with drift.
\end{itemize}

These results can easily be shown to apply as well to general
$n$-state models. Moreover, it can be argued that even if several
parallel channels exist for moving from one monomer to another the
results are also qualitatively unchanged \cite{Kafri04}.

Experimentally, the predictions of the model can be tested by
measuring the displacement of the motor as a function of time,
averaged over different experimental runs (and, possibly,
sequences). Each time trace of the motor position will have an
irregular shape due to pauses, which will increase in duration as
the stall force is approached, at specific locations along the
track. However, averaged over many time traces (or sequences) the
expected displacement will grows as $t^\mu$ with $\mu<1$ close
enough to the stall force. Note that if one averages over time
traces for a fixed sequence, the displacement is expected to grow as
$s(t) t^\mu$ where $s(t)$ has fluctuations of order unity, because
the sequence information is not completely erased in this case. An
alternative experimental test would be to measure the distribution
of dwell times ${\cal P}(\tau)$. Because ${\cal P}(\tau) \sim
1/\tau^{1+\mu}$ for large $\tau$, the distribution becomes {\it
wider} as the stall force is approached. Monitoring ${\cal P}(\tau)$
has the advantage of probing the wide distribution even in regimes
which are not very close to the stall force.

We now consider limiting cases of the above model on heterogeneous
tracks. These illustrate a number of interesting features and
suggest ways in which the anomalous dynamics might be observed
experimentally. We will also use the model to explore heterogeneous
chemical energy sources for motors on a periodic track. This
situation may be realized in motors such as kinesin which can use
several types of chemical energy to move along the track. From the
two state model described above we deduce the expected behavior of
the velocity as the relative proportions of the different fuels is
varied and mention generalizations to more general $n$-state models.

\section{Strongly biased motors}

In this section we study motors moving on a {\it heterogeneous
substrate} in the limit where one of the transition rates is
strongly biased in a certain direction. An extreme limit occurs when
one of the transition rates, in, say, the backward direction, is
zero. Although this limit violates detailed balance, it could be a
reasonable approximation for certain strongly biased experiments.
One such model is a special case of the two state model discussed in
Section \ref{secmodel}:
\begin{eqnarray}
w_a^\rightarrow&=&(\alpha e^{\Delta \mu /T} + \omega)e^{-\Delta
\varepsilon/T-f/2T}
 \nonumber \\
w_b^\leftarrow&=&(\alpha  + \omega)e^{f/2T}
 \nonumber \\
w_a^\leftarrow &=& \omega' \; e^{-\Delta \varepsilon/T+f/2T}
\label{eq:rates2}
\\
w_b^\rightarrow&=& \omega' \; e^{-f/2T} \;.  \nonumber
\end{eqnarray}
Here we have set $\alpha'=0$ so that in the $\Delta \mu \gg T$ limit
the motor will be strongly biased to move towards the right.
Physically, this situation corresponds to a motor which can use
chemical energy only to move in a certain direction. (We expect
qualitatively similar results for a wide variety of strongly biased
``one way'' models). Next, we assume an extremely strong bias
$\Delta \mu \gg T$ limit of the model such that $\alpha e^{\Delta
\mu /T}$ is very large but $\alpha$ and $\omega$ are so small that
$w_b^\leftarrow$ can be set to be zero. This limit only makes sense
far from the stall force. The stall force of a model with
$w_b^\leftarrow=0$ (as any other model were one of the reactions is
assumed to be unidirectional) is infinite: The effective energy
landscape, Eq. \ref{eq:dE}, which describes the dynamics of such a
limit has an infinite slope. In this section we will compare
numerically trajectories when $w_b^\leftarrow=0$ and when
$w_b^\leftarrow \neq 0$.

\begin{figure}
\centering
\includegraphics[width=12cm]{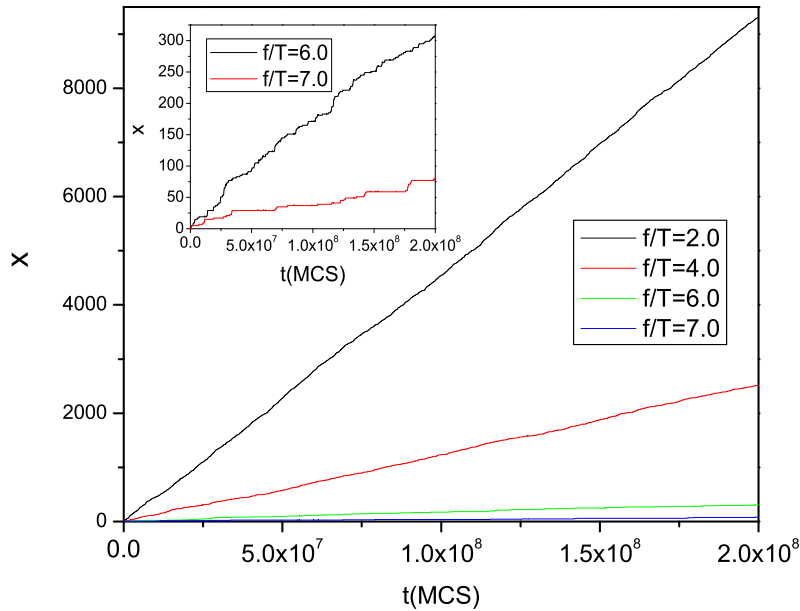} \caption{Sample trajectories obtained on a heterogeneous
track using the model with $w_b^\leftarrow=0$ for various values of
the reduced force $f/T$. The lower the trajectory the higher the
opposing force. Here we took with equal probability
$\alpha=5,\omega=1,\omega'=2$ and $\alpha=0.2,\omega=1,\omega'=1$.
In both cases $\Delta \varepsilon=0$ and $ e^{\Delta \mu/T}=500$. In
the inset the two largest values of the force are presneted in more
detail. \label{fig:noback}}
\end{figure}

Using Eq. \ref{eq:mu}, the infinite slope of the effective energy
landscape implies that $\mu=\infty>2$, so that the dynamics is
diffusion with drift. The linear drift is illustrated clearly in
Fig. \ref{fig:noback} where trajectories of the strongly biased
model are shown for different forces. Note that even for very large
forces the displacement of the motor as a function of time grows
linearly. It can be shown that the dwell time distribution on the
track decays exponentially ${\cal P}(\tau) \sim e^{-\tau/\tau^*}$ in
contrast to the power law distribution expected when $w_b^\leftarrow
\neq 0$. To observe any significant pausing clearly one must have $f
\simeq \Delta \mu$ (see inset of Fig. \ref{fig:noback}). In this
limit, however, $w_a^\leftarrow \gtrsim w_a^\rightarrow$, and
backwards motion cannot be neglected (see Eq. \ref{eq:rates2}).

\begin{figure}
\centering
\includegraphics[width=12cm]{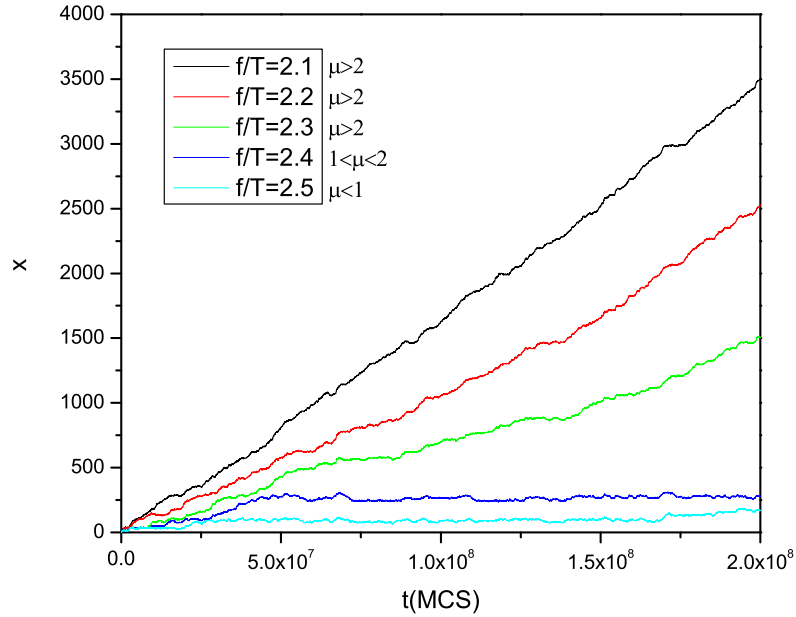} \caption{Trajectories obtained using the
model of Eq. \ref{eq:rates2} with
$w_b^\leftarrow=(\alpha+\omega)e^{f/2T}$ and
$w_a^\rightarrow=(\alpha e^{\Delta \mu /T}+\omega)e^{-\Delta
\varepsilon /T-f/2T}$.  The lower the trajectory the higher the
resisting force. Here we took with equal probability
$\alpha=5,\omega=1,\omega'=2$ and $\alpha=0.2,\omega=1,\omega'=1$
along the heterogeneous track. In both cases $\Delta \varepsilon=0$
and $ e^{\Delta \mu/T}=500$. Anomalous displacement occur for
$f/T>2.47$ while anomalous diffusion occurs for $f/T>2.38$. The
stall force is $f_s=2.62$. \label{fig:back}}
\end{figure}

In Fig. \ref{fig:back} we show the very different behavior of
simulations where we take the back hopping to be non-zero. Now we do
not neglect $\omega$ and the backward hopping rate $w_b^\leftarrow$
is nonzero! For forces near the stall force the displacement of the
motor as a function of time seems to saturate even for a trajectory
generated by a {\it single} numerical experiment for a particular
sequence. Such a behavior is consistent with that expected from a
sublinear displacement of the motor. Note, however, that with the
exception of the largest force, all curves on long enough time
scales are expected to yield, after an average over many thermal
realizations, an asymptotically linear curve (see the corresponding
values of $\mu$ presented in the figure). However, even for these
curves, pausing is pronounced despite the fact that for small
resisting forces the motor rarely moves backwards.

\section{Effect of heterogeneous fuel concentrations for a motor moving along a
heterogeneous substrate}

For some molecular motors the type of fuel which is used to move
depends on the specific site along the track. For example, in the
case of RNA polymerase, which produces messenger RNA, the energy
from the hydrolysis of the specific NTP which is added to the mRNA
chain is used for motion. While a random forcing energy landscape
would exist even if the chemical energy released from every NTP were
the same (see Sec. 2), the different chemical energies enhance the
variance of the slopes, $V$, of the random forcing energy landscape
(due to the different chemical potentials of the nucleotides in the
transcript). This in turn (see Eq. \ref{eq:mu}) lowers the value of
$\mu$ as compared to the case of equal chemical energies.

As suggested in \cite{Kafri04} the variance $V$ could be further
increased by increasing the concentration difference between the
different NTPs in the solution. In an extreme situation, where one
of the NTPs is completely removed from the solution, the motor will
stall at specific locations where the NTP is needed. This trick is
used in experiments to synchronize and control the motion of RNA
polymerases \cite{trick}. In this section we illustrate using the
simple model, Eq. \ref{eq:rates}, the effects of changing NTP
concentrations on motor motion. Although we work within a ``minimal
model'' we expect the same effects for more complicated models of
molecular motors with many internal states (for an example of such a
model for RNA polymerase see \cite{Lucy}).

\begin{figure}
\centering
\includegraphics[width=12cm]{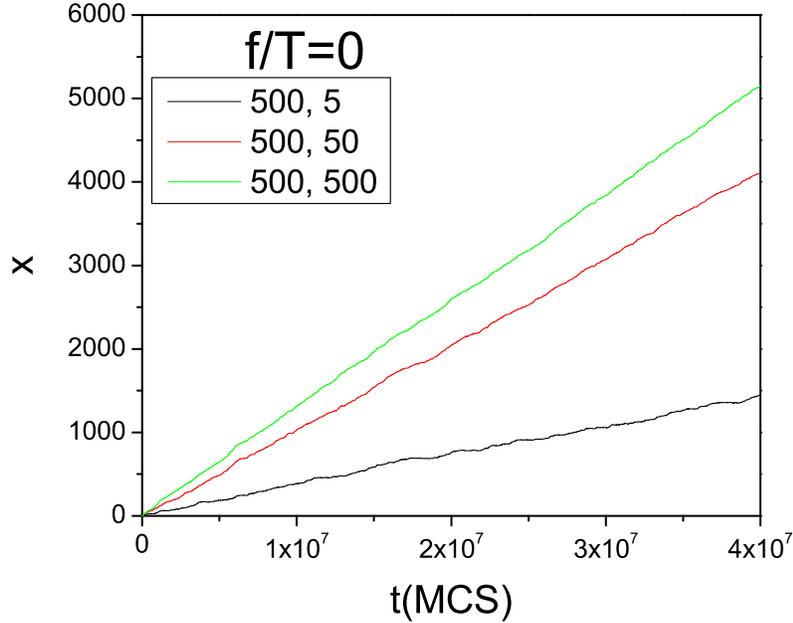} \caption{Motor trajectories obtained using the
model defined by Eq. (\ref{eq:rates2}). Here, substrate
heterogeneity is incorporated by taking with equal
 probability $\{ \alpha=5,\omega=1,\omega'=2 \}$ and
 $\{ \alpha=0.2,\omega=1,\omega'=1 \}$.
The corresponding values of $e^{\Delta \mu/T}$ for the two different
fuels are denoted in the box. In both cases $\Delta \varepsilon/T
=0$ and $f/T=0$. Lower trajectories correspond to a larger
difference in the chemical potentials associated with the fuels.
\label{fig:f0}}
\end{figure}

To this end, we studied motors which can use two kinds of fuels
depending on its location along the track. We hold the concentration
of one fuel fixed and lower the concentration of the other by
reducing the chemical potential associated with it. For reference we
also study the case when the fuels have equal chemical potentials.
Fig. \ref{fig:f0} displays results of numerical simulations of the
model defined by Eq. (\ref{eq:rates2}). Similar results were
obtained with the more general model of Eq. (\ref{eq:rates}). We
show simulations with $f/T=0$ and when for one fuel $e^{\Delta
\mu/T}=500$ and for the other $e^{\Delta \mu/T}=500,50$ or $5$. From
a generalization of Eq. \ref{eq:dmu}, we see that this latter
variation corresponds to a change of two orders of magnitude in the
concentration of the second fuel \cite{HowardBook}. As can be seen
from Fig. \ref{fig:f0}, the effect of reducing one of the chemical
potentials is to slow down the motion. However, note that for all
fuel concentrations the motor displacement as a function of time is
linear with these parameter values. Since at zero applied force one
expects the motor always to be biased preferentially in a given
direction which is independent of the monomer this behavior will
hold even for larger differences in concentration.

\begin{figure}
\centering
\includegraphics[width=12cm]{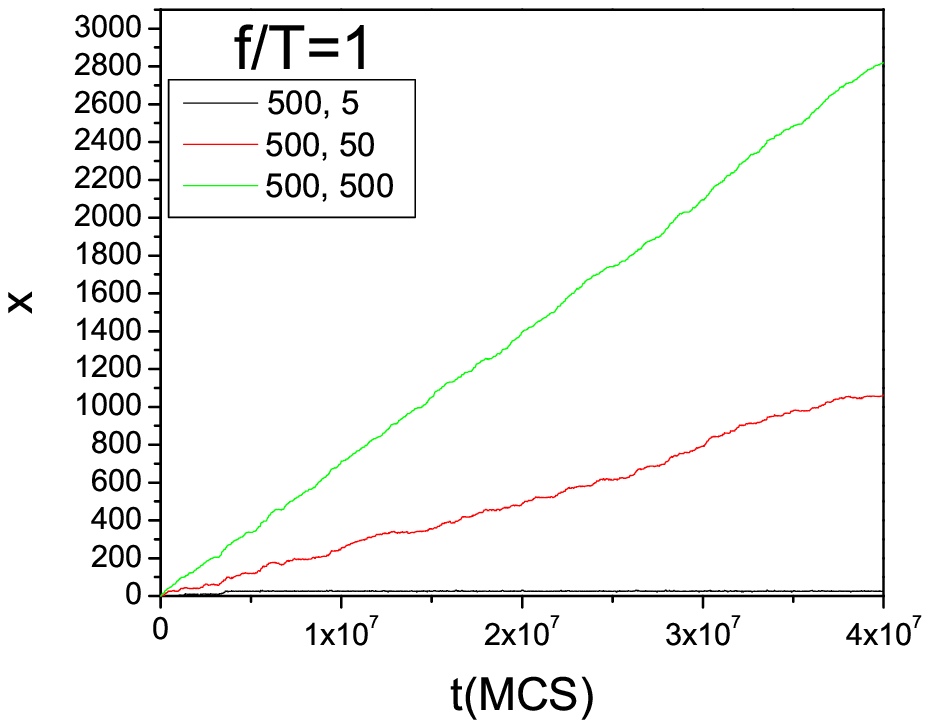} \caption{Motor trajectories
obtained using the model Eq. (\ref{eq:rates2}) with an opposing
force. We again took with equal probability $\{
\alpha=5,\omega=1,\omega'=2 \}$ and $\{
\alpha=0.2,\omega=1,\omega'=1 \}$, with the corresponding values of
$e^{\Delta \mu/T}$ for the two different fuels as indicated in the
figure. In both cases $\Delta \varepsilon/T=0$ and $f/T=1$. Lower
trajectories correspond to a larger difference in the chemical
potential differences. \label{fig:f1}}
\end{figure}

The change in concentration of one of the fuels can make the regime
of anomalous dynamics much larger. In Fig. \ref{fig:f1} we show the
result of applying a force which opposes the motion of the motors on
the dynamics. As can be easily seen from the figure, the velocity of
the motors is, of course, lower in comparison to that when no force
is applied. However, note that the motion of the curve with the
largest difference in chemical potential (barely visible at the
bottom) is very different. The motor almost immediately stalls after
it starts moving. This example illustrates clearly how changing the
concentrations of the different fuels can make the regime of
anomalous dynamics easy to access at lower forces. In fact, averaged
over many thermal and sequence realizations, the curve with the
largest difference in chemical potentials shows a displacement of
the motor which grows sublinearly in time, indicating that $\mu<1$.

\section{Heterogeneous fuels on periodic tracks}

In this section we consider a molecular motor which is moving along
a {\it periodic track} in a solution which contains more than one
type of molecule which can supply it with chemical energy. The
situation arises for kinesin in the presence of both ATP and, say,
GTP. It is known that, while less efficient, alternative NTP
molecules can also be used by kinesin to move along the track
\cite{Kin1,Kin2,Kin3}. We consider, for simplicity, a solution with
two types of chemical fuels within the simple two state model (The
analysis presented can easily be generalized to include additional
fuel types). In addition, we generalize our model to treat two
separate cases. In the first we assume that the internal states of
the motor (i.e., the $a$ and $b$ sites in Fig. \ref{fig:motor}) are
independent of the fuel used so that the chemical fuels are only
used to move between states with fuel-dependent potential
differences driving the changes. This situation arises when the fuel
from the motor is used and released so quickly that the motor is
unbound to the fuel in the ``excited'' internal state. In this case
the fuel binding to the motor does not define an internal state. In
the second case we study we allow for additional internal states of
the motor, depending on which type of chemical fuel is bound to it.
This situation arises when internal ``excited'' states include a
fuel bound to the motor. Since the motor bound to the two fuels
defines two distinct internal states, a direct thermal transition
between them is therefore not possible. While both cases discussed
below involve parallel pathways for transitions across a monomer,
they are distinct in the type of internal states.

\subsection{Case I: Internal states of the motors are independent of
the chemical fuel}

This case amounts to a straightforward generalization of the rates
in Eq. \ref{eq:rates} to allow for multiple fuels. As mentioned
above, we assume the internal states of the motor are independent of
the chemical energy and that chemical energy is only used to assist
the transitions. With two different chemical fuels, the generalized
rates entering Eqs. \ref{eq:b} and Eq. \ref{eq:a} are now
\begin{eqnarray}
w_a^\rightarrow&=&(\alpha_1 e^{\Delta \mu_1 /T} + \alpha_2 e^{\Delta
\mu_2 /T}+ \omega)e^{-\Delta \varepsilon/T-f/2T}
 \nonumber \\
w_b^\leftarrow&=&(\alpha_1 + \alpha_2  + \omega)e^{f/2T}
 \nonumber \\
w_a^\leftarrow &=&(\alpha_1' e^{\Delta \mu_1 /T} +  \alpha_2'
e^{\Delta \mu_2 /T} +\omega')e^{-\Delta \varepsilon/T+f/2T}
\label{eq:ratestwoB}
\\
w_b^\rightarrow&=&(\alpha_1' + \alpha_2'  + \omega')e^{-f/2T} \;.
\nonumber
\end{eqnarray}
There are now three parallel paths, two assisted by chemical energy,
one thermal. The subscript $1$ or $2$ refers to whether fuel ``$1$''
or ``$2$'' is being used for the transition. Standard relations for
the chemical potential difference \cite{HowardBook} imply that
$e^{\Delta \mu_i/T}$ grows linearly with the concentration of fuel
``$i$'' (see Eq. \ref{eq:dmu}). Because the extra channel is
present, when one of the chemical potential differences is set to
zero the model reduces to the single motor model but with modified
rates as compared to a motor in a solution where the fuel is
completely absent. For example, if $\Delta \mu_2=0$ the model can be
written in terms of Eq. \ref{eq:rates} with the identifications
$\omega \rightarrow \omega+\alpha_2$ and $\omega' \rightarrow
\omega'+\alpha'_2$. Note also that, in contrast to disordered tracks
where the motion of the motor is described by a random walker moving
on a random force landscape, here the energy landscape is periodic
except for a well defined tilt given by Eq. \ref{eq:dE}.

\begin{figure}
\centering
\includegraphics[width=12cm]{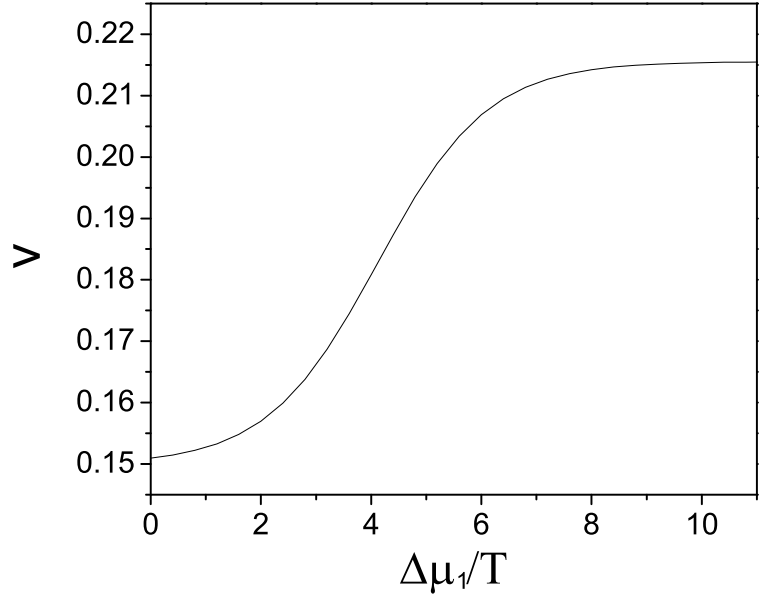} \caption{The velocity as a function of the chemical potential
difference $\Delta \mu_1/T$ of fuel ``$1$''. The velocity is plotted
in units where is size of a monomer (i.e., the lattice constant
$2a_0$ in Fig. \ref{fig:motor}) is set to be one and arbitrary time
units. We used a ``nearly one way model''  based on Eq.
\ref{eq:ratestwoB} where the rates were chosen to be
$\alpha_1=5,\alpha_2=2,\alpha_1'=0.01,\alpha_2'=0.02,\omega=0.01,\omega'=0.2$
in arbitrary units of inverse time and $\Delta \mu_2/T=5,f=0,\Delta
\varepsilon =0 $. For reference the velocity of with only fuel of
type ``2'' with the same value of $\Delta \mu_2/T$ is
$v=0.195$.\label{fig:fuel1}}
\end{figure}

\begin{figure}
\centering
\includegraphics[width=12cm]{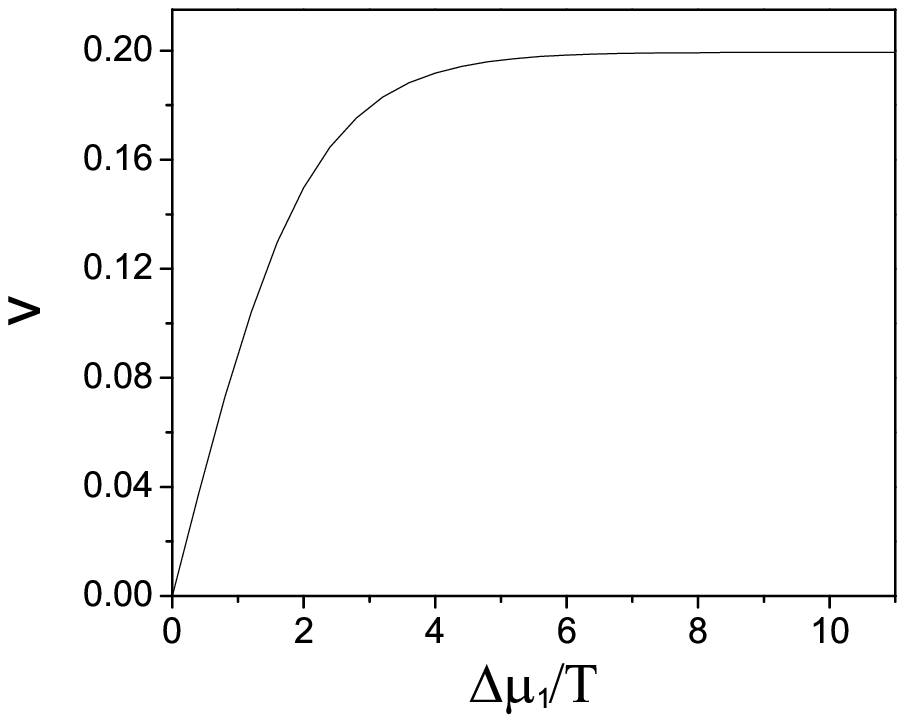} \caption{The velocity in arbitrary time units
as a function of the chemical potential difference $\Delta \mu_1/T$
where the other chemical channel associated with fuel ``2'' is
closed. The velocity is plotted in units where is size of a monomer
is set to be one. The rates were chosen to be
$\alpha_1=5,\alpha_2=0,\alpha_1'=0.01,\alpha_2'=0.0,\omega=0.01,\omega'=0.2,f=0$
in arbitrary units of inverse time and $\Delta \varepsilon =0$.
\label{fig:fuel2}}
\end{figure}

With the help of  Eq. \ref{eq:vel}, it is straightforward to verify
that for this model the velocity takes the form
\begin{equation}
v=\frac{A+B[x_1]+B'[x_2]}{C+D[x_1]+D'[x_2]} \;, \label{eq:ratio}
\end{equation}
where $[x_1]$ and $[x_2]$ are the concentrations of fuel $1$ and $2$
respectively and $A,B,B',C,D,D'$ are complicated functions of the
the coefficients in Eq. \ref{eq:ratestwoB}. The velocity thus varies
as ratio of two polynomials, each depending linearly on the
concentration of both fuels. Consider, for example, the velocity in
an experiment where one of the fuels is held at constant
concentration and the concentration of the other is varied. The
external force is set to $f=0$; (varying $f$ did not affect the
qualitative features discussed below). As shown in Fig.
\ref{fig:fuel1}, the average velocity smoothly crosses over from a
small value to a larger value in a sigmoidal fashion as the
concentration of more energy rich fuel (``fuel $1$'') is increased.
Note that $\Delta \mu_1/T$ varies {\it logarithmically} with the
concentration of fuel number $1$ \cite{HowardBook}. Thus, the
concentration of fuel ``$1$'' varies over many orders of magnitude
for the range of $\Delta \mu_1/T$ shown in Fig. \ref{fig:fuel1}. In
typical experiments, only a small portion of this crossover may be
visible. For small $\Delta \mu_1/T$, motor movement is controlled by
fuel ``$2$'' while for large $\Delta \mu_1/T$ it is controlled by
fuel ``$1$''. An analogous plot for motor velocity vs. $\Delta
\mu_1/T$ when fuel $2$ is absent entirely is shown for reference in
Fig. \ref{fig:fuel2}. As expected, the velocity no longer exhibits a
sigmoidal crossover between two regimes.


\subsection{Case II: Internal states of the motors are coupled to
the chemical fuel}

Next, we consider a different model incorporating two fuels. We now
assume that the type of fuel molecule used to move the motor
determines the {\it entire} chemical cycle which leads the motor to
move across one monomer. An example might be the $n=2$ motor
landscape shown in Fig. \ref{fig:motor} where one step (e.g.; $a$
site $\rightarrow$ $b$ site) involves the fuel molecule binding to
the motor. In this case the entire sequence of transitions would be
dictated by the fuel that is used. Thus, the choice of rates (either
in the forward or backward direction) would be dictated by the
initial step which is chosen at random, depending on the type of
fuel utilized.

Since the state of the motor is directly coupled to the chemical
fuel, a pure thermal transition between the states is not possible
(unless it involves moving across the monomer through parallel
pathways not related to the motor, an effect which is ignored here).
The new feature is that the motor can move across a monomer by
choosing one of two distinct chemical pathways. This is in contrast
to the model defined by Eq. \ref{eq:ratestwoB} where two distinct
chemical pathways exist for passing between the internal states of
the motor, but where the transition across the monomer can occur via
a mixture different chemical (or thermal) pathways.

An example for such a model, where each fuel is modeled by a
distinct channel which is a special case of the two state model
defined by (\ref{eq:rates}) is given by:
\begin{eqnarray}
w_a^\rightarrow&=&\alpha e^{\Delta \mu_1 /T-f/2T}
 \nonumber \\
w_b^\leftarrow&=&\alpha e^{f/2T}
 \nonumber \\
w_a^\leftarrow &=& \omega \; e^{f/2T} \label{eq:parrates}
\\
w_b^\rightarrow&=& \omega \; e^{-f/2T} \;,  \nonumber
\end{eqnarray}
for the first channel and
\begin{eqnarray}
u_a^\rightarrow&=&\gamma e^{\Delta \mu_2 /T-f/2T}
 \nonumber \\
u_b^\leftarrow&=&\gamma e^{f/2T}
 \nonumber \\
u_a^\leftarrow &=& \nu \; e^{f/2T} \label{eq:parrates2}
\\
u_b^\rightarrow&=& \nu \; e^{-f/2T} \;.  \nonumber
\end{eqnarray}
for the second channel. The fuel concentrations enter through the
chemical potential differences $\Delta \mu_1$ and $\Delta \mu_2$
(see Eq. \ref{eq:dmu}). The relative fuel abundances therefore
control the ratio of the rates $w_a^\rightarrow$ and
$u_a^\rightarrow$. The model will be realized physically in a motor
which binds a fuel and uses its chemical energy in the transitions
$w_a^\rightarrow$ or $u_a^\rightarrow$. The transitions
$w_b^\rightarrow$ and $u_b^\rightarrow$ involve the release of the
corresponding fuel. Here for simplicity we have set the energy
difference between the states to be zero and neglected parallel
thermal channels. We do not expect the qualitative results described
below to be affected by such complications.

The velocity of the model can be calculated in a straightforward
manner and is found to be
\begin{equation}
v=\frac{(u_b^\rightarrow+u_b^\leftarrow)(w_a^\rightarrow
w_b^\rightarrow - w_a^\leftarrow
w_b^\leftarrow)+(w_b^\rightarrow+w_b^\leftarrow)(u_a^\rightarrow
u_b^\rightarrow - u_a^\leftarrow
u_b^\leftarrow)}{(u_b^\rightarrow+u_b^\leftarrow)(w_b^\rightarrow+w_b^\leftarrow)+
(u_b^\rightarrow+u_b^\leftarrow)(w_a^\rightarrow+w_a^\leftarrow)+
(u_a^\rightarrow+u_a^\leftarrow)(w_b^\rightarrow+w_b^\leftarrow)}
\;.
 \label{eq:veltwochan}
\end{equation}
Note that as in Case I, the velocity behaves as a ratio of two
polynomials which are linear in the $e^{\Delta \mu_i /T}$, and hence
with each of the fuel concentrations (similar to Eq.
\ref{eq:ratio}). Also similar is the fact that the presence of a
second channel alters the velocity even when one of the chemical
potential difference is zero. Again, due to the presence of the
second channel the velocity is lower than that of the motor in the
presence of a single fuel. We do not present plots of the resulting
velocity as the concentration of one of the fuels is varied since
the qualitative features are similar to those presented in the
previous subsection.

Finally, we comment that more complicated scenarios (for example,
the existence of additional parallel pathways, or more general
$n$-state models) might change the explicit dependence on the
concentrations of the fuels. However, we expect a general form of a
ratio of two polynomials in the concentrations of the fuels even in
much more complicated scenarios. Indeed, with a specific experiment
in mind and some structural information on the motor one might be
able to use such experiments, coupled with an analysis similar to
that presented above, to deduce the number of steps in the chemical
cycle which depend explicitly on the fuel concentration. For
example, we have considered models where the internal states are
independent of the chemical fuel and found that the general velocity
can be a ratio of polynomials of a degree which is related to the
number of steps which depend on the chemical fuel. However, the
detailed results were very dependent on the choice of rates made.

\noindent {\bf Acknowledgments:} We are very grateful to D. K.
Lubensky for many stimulating conversations and the collaboration
which led to references \cite{Kafri04} and \cite{KLN05}. We also
thank L. Bau, M. D. Wang and K. C. Neuman for useful conversations
and J. Gelles and N. Guydosh for interesting us in the problem of
different fuels for kinesin on periodic tracks like microtubules.
D.R.N was supported by the National Science Foundation through Grant
No. DMR-0231631 and the Harvard Materials Research Laboratory via
Grant No. DMR-0213805. Y.K was supported by the Human Frontiers
Science Program.

\vspace{0.5cm}

\noindent $^*$ Permanent address: Department of Physics, Technion,
Haifa 32000, Israel.

\end{document}